\documentclass[letterpaper]{article} % DO NOT CHANGE THIS
\usepackage{aaai23}  % DO NOT CHANGE THIS
\usepackage{times}  % DO NOT CHANGE THIS
\usepackage{helvet}  % DO NOT CHANGE THIS
\usepackage{courier}  % DO NOT CHANGE THIS
\usepackage[hyphens]{url}  % DO NOT CHANGE THIS
\usepackage{graphicx} % DO NOT CHANGE THIS
\usepackage{natbib}  % DO NOT CHANGE THIS AND DO NOT ADD ANY OPTIONS TO IT
\usepackage{caption} % DO NOT CHANGE THIS AND DO NOT ADD ANY OPTIONS TO IT
\usepackage{algorithm}
\usepackage{algorithmic}

\usepackage{newfloat}
\usepackage{listings}
\DeclareCaptionStyle{ruled}{labelfont=normalfont,labelsep=colon,strut=off} % DO NOT CHANGE THIS
\floatstyle{ruled}
\newfloat{listing}{tb}{lst}{}
\floatname{listing}{Listing}
\title{Periodic Multi-Agent Path Planning}
\author {
    Kazumi Kasaura,
    Ryo Yonetani,
    Mai Nishimura
}
\affiliations {
    OMRON SINIC X Corporation
    Hongo 5-24-5, Bunkyo-ku, Tokyo, Japan\\
    kazumi.kasaura@sinicx.com, ryo.yonetani@sinicx.com, mai.nishimura@sinicx.com
}
\def\eg{{\it e.g.}}

\def\ie{{\it i.e.}}

\usepackage{tabularx}
\usepackage{booktabs}

\usepackage{amsmath, amsfonts, subfigure}

\newtheorem{proposition}{Proposition}

\begin{document}

\maketitle

\begin{abstract}
Multi-agent path planning (MAPP) is the problem of planning collision-free trajectories from start to goal locations for a team of agents. This work explores a relatively unexplored setting of MAPP where \emph{streams} of agents have to go through the starts and goals with high throughput. We tackle this problem by formulating a new variant of MAPP called periodic MAPP in which the timing of agent appearances is periodic. The objective with periodic MAPP is to find a periodic plan, a set of collision-free trajectories that the agent streams can use repeatedly over periods, with periods that are as small as possible. To meet this objective, we propose a solution method that is based on constraint relaxation and optimization. We show that the periodic plans once found can be used for a more practical case in which agents in a stream can appear at random times. We confirm the effectiveness of our method compared with baseline methods in terms of throughput in several scenarios that abstract autonomous intersection management tasks.
\end{abstract}

\section{Introduction}

Multi-agent path planning (MAPP) refers to the problem of finding a set of collision-free trajectories from start to goal locations for a team of multiple agents. MAPP, specifically multi-agent pathfinding (MAPF) that searches for a solution on a given graph, is a fundamental problem in multi-agent systems~\cite{stern2019multi}.

We are particularly interested in the relatively unexplored problem of MAPP in which, rather than a single agent, \emph{a stream of agents} enters each start location and leaves the environment upon reaching the goal. Instead of finding a set of feasible trajectories with a small total cost, we aim to improve the throughput for agent streams passing through the environment. Such settings would be beneficial for several practical applications, such as autonomous intersection management (AIM)~\cite{dresner2008multiagent} and automated warehouses~\cite{wurman2008coordinating}.

Handling agent streams in such a problem setting poses a nontrivial technical challenge. As the throughput increases, the environment would be filled with a large number of agents, making it difficult to use optimal planning algorithms with limited scalability (\eg, conflict-based search~\cite{sharon2015conflict}). It is also not obvious if the high throughput can be maintained with scalable planners that nevertheless have to determine agent trajectories adaptively in sequence (\eg, prioritized planning~\cite{silver2005cooperative}). Furthermore, finding collision-free trajectories in highly-crowded environments would require consideration of planning in the continuous space (\ie, not grid maps) and with continuous time (\ie, allowing agents to start and stop at an arbitrary timing in a continuous timeline). However, such continuous setups are generally challenging and there are few established approaches~\cite{andreychuk2021improving,andreychuk2022multi,kasaura2022prioritized}.

In this work, we start by formulating a bit simplified but new variant of MAPP called \emph{periodic multi-agent path planning (periodic MAPP)} in which the timing of agent appearances is periodic. The objective with periodic MAPP is to find a \emph{periodic plan}, \ie, a set of collision-free trajectories that streams of agents can use repeatedly over periods. By finding such plans with periods that are as small as possible, we are able to improve the throughput of agent streams.
Importantly, periodic plans once found can be easily used for solving a more practical problem called online MAPP, a variant of online MAPF~\cite{vsvancara2019online} in which a stream of agents can appear at random times but can also wait until the subsequent agents enter the environment.

We develop a constraint-relaxation and optimization method as a solution method to periodic MAPP. With this method, we first generate an initial periodic plan under relaxed constraints about the physical size of agents with an arbitrarily large period they appear. We then solve a continuous optimization problem to improve the initial plan such that the agent size matches the original one and the period becomes as small as possible. Therefore, our method can find a collision-free and repeatable plan while minimizing the period. We provide insights into the topological aspect of solutions obtained with the proposed method.

\looseness=-1
We evaluated the effectiveness of our method on several scenarios of abstracting AIM tasks, in which the goal is to move vehicles appearing at intersections to the other side without collision. Unlike existing methods that require planning or re-planning for every appearance of a new vehicle, the proposed method using periodic plans does not necessitate communications with other vehicles to retrieve their current locations or to update their trajectories. Nevertheless, the solutions derived from the proposed method are comparably good or sometimes even better in terms of the throughput, compared with those from baseline methods that combine online MAPP algorithms~\cite{vsvancara2019online} and MAPP algorithms for continuous spaces and times~\cite{andreychuk2022multi,andreychuk2021improving,kasaura2022prioritized}.

\section{Periodic MAPP}\label{sec:problem}

\begin{figure}[t]
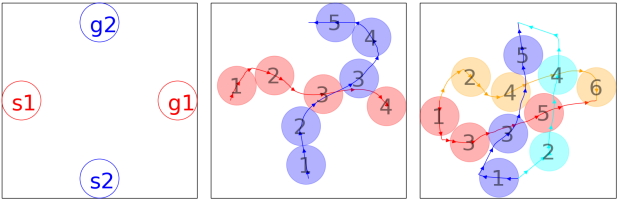

    \centering
    \includegraphics[width=0.32\linewidth]{figures/problem.pdf}
    \includegraphics[width=0.32\linewidth]{figures/plan_1.pdf}
    \includegraphics[width=0.32\linewidth]{figures/plan_2.pdf}
    \caption{Example of periodic MAPP problem with $N=2$ (left) and periodic plans with $M=1$ (middle) and $M=2$ (right). Solid lines show trajectories. Numbered circles indicate where agents are at each elapsed period after their appearance.}

    \label{fig:teaser}
\end{figure}

\paragraph{Overview.}
We consider a two-dimensional (2D) environment with several pairs of start and goal locations. For each start location, a stream of agents appears periodically with a user-defined \emph{period} (\ie, time interval). Each agent must move to its goal while avoiding collisions with the borders of the environment and other agents and disappear from the environment upon reaching the goal. We assume that there exists a certain \emph{cycle}, the number of periods within which we can find a \emph{periodic plan} (\ie, a set of collision-free trajectories that can be used periodically over cycles). Therefore, a periodic plan may span multiple periods as collision-free trajectories for agents appearing at the same locations that are not necessarily the same across periods (see agents shown in red/orange or those in blue/cyan in Fig.~\ref{fig:teaser}.) Informally, the objective with periodic MAPP is to find such periodic plans for a given cycle with periods that are as small as possible. In other words, we wish to produce a high throughput plan that enables us to move agents from their respective start to the goal even if they appear in rapid succession. Note that, if the non-periodic version of a given problem instance (\ie, standard MAPP with a single agent appearing from each start) has a solution, the problem instance has a periodic plan for any cycle with the period given by the arrival time of the last agent. For simplicity, we assume that all agents have bodies modeled by circles with the same fixed radius and follow a simple kinodynamic constraint in which the velocity cannot exceed a certain maximum limit. 

\paragraph{Problem instances.}
Formally, we model a problem instance of periodic MAPP by a tuple $(\mathcal{E}, \{(s_1,g_1),\dots,(s_N, g_N)\}, r, v_\mathrm{max})$, where $\mathcal{E}\subseteq \mathbb{R}^2$ is a set of valid states in the 2D environment. The set $\{(s_1,g_1),\dots,(s_N, g_N)\}$, where $s_n, g_n\in\mathcal{E}$, describes $N$ pairs of start and goal locations for agent streams. The variables $r$ and $v_\mathrm{max}$ are the radius and maximum velocity of each agent, respectively. 
\paragraph{Periodic plans.}
We refer to a \emph{period} as $\tau\in\mathbb{R}_+$, a time interval with which a new set of agents can appear at respective start locations $s_1,\dots,s_N$. While denoting $\emph{cycle}$ as $M\in\mathbb{N}_+$, we describe a \emph{periodic plan with $M$ periods} by a set of $M\times N$ trajectories $\Gamma_M=(\gamma_{n,m}: [0, T_{n,m}] \rightarrow \mathcal{E})_{1\leq n \leq N, 0\leq m < M}$. The periodic plan should satisfy the following conditions:
\begin{itemize}
\item (Start and goal locations) For all $1\leq n\leq N$, $0\leq m < M$,
 $\gamma_{n,m}(0) = s_n$, and $\gamma_{n,m}(T_{n,m}) = g_n$.
\item (Maximum velocity) For all $1\leq n\leq N$, $0\leq m<M$, and $t\in [0, T_{n,m}]$, the velocity of agents satisfies
\begin{equation}
    \left|\frac{d\gamma_{n,m}}{dt}(t)\right| \leq v_{\mathrm{max}}.
    \label{eq:vel}
\end{equation}
\item (Clearance from boundaries) Let $\mathrm{dist}_{\mathcal{E}}(x)$ be the distance of $x\in \mathcal{E}$ from the boundary of $\mathcal{E}$.
Then, for all $1\leq n\leq N$, $0\leq m<M$, and $t\in [0, T_{n,m}]$,
\begin{equation}
\mathrm{dist}_{\mathcal{E}}(\gamma_{n,m}(t))\geq r.
    \label{eq:obs}
\end{equation}
\item (Collision-free)\footnote{The agent appearing at $s_n$ at time $(aM+m)\tau$ follows the trajectory $\gamma_{n,m}$ for any $a\in\mathbb{Z}$. Since there exists an agent at $\gamma_{n,m}(t)$ when the time is $\ldots, (-2M+m)p\tau + t, (-M+m)\tau + t, m\tau + t, (M+m)\tau+t,(2M+m)\tau + t,\ldots$, there exist two agents at $\gamma_{n,m}(t)$ and $\gamma_{n',m'}(t')$ at the same time if $(m-m')\tau+(t-t')\in M\tau\mathbb{Z}$, except when $(n,m,t)=(n',m',t')$. To avoid a collision between them, their distance must be at least $2r$.}
For all $1\leq n, n'\leq N$, $0\leq m, m'<M$, and $t\in [0, T_{n,m}]$, $t'\in [0, T_{n',m'}]$ such that $(m-m')\tau+(t-t')\in M\tau\mathbb{Z}$ and $(n,m,t)\neq (n',m', t')$,
\begin{equation}\label{eq:collision-free}
\left|\gamma_{n,m}(t)-\gamma_{n',m'}(t')\right|\geq 2r.
\end{equation}
\end{itemize}

\paragraph{Objective of periodic MAPP.}
Given a problem instance $(\mathcal{E}, \{(s_1,g_1),\dots,(s_N, g_N)\}, r, v_\mathrm{max})$ and a cycle $M$, our objective is to find periodic plans $\Gamma_M$ with periods $\tau$ that are as small as possible.

\section{Solution Method}\label{sec:method}

\looseness=-1
In this section, we explain the proposed solution method for producing periodic plans for periodic MAPP. Specifically, we use a two-step approach that first derives initial solution trajectories for a relaxed problem that ignores the constraints about $r$ and the objective for $\tau$. We then optimize them by solving a continuous optimization problem to satisfy all the original conditions and improve the solution quality with respect to $\tau$. This is a reasonable approach to derive solution trajectories in a continuous space and time setup while aiming to minimize the continuous period value that affects the solution.

\subsection{Trajectory Representation}
We represent each trajectory $\gamma_{n, m}$ by a sequence of $K+1$ timed locations with a timestep $\Delta t_{n, m}$, \ie, $((x_{n, m, 0}, 0), (x_{n, m, 1}, \Delta t_{n, m}), \ldots, (x_{n, m, K}, K\Delta t_{n, m}))$ where $x_{n, m, k} \in \mathcal{E}$, $x_{n, m, 0} =s_n$, $x_{n, m, K}=g_n$, and $K\Delta t_{n, m} = T_{n, m}$. Agents are assumed to move between two locations $x_{n,m,k}, x_{n,m,k+1}$ in a straight line with constant velocity.
This discretized representation should also satisfy the maximum velocity condition in Eq.~(\ref{eq:vel}) that is rewritten as
\begin{equation}
v_{n,m,k}:=\left|\frac{x_{n,m,k+1}-x_{n,m,k}}{\Delta t_{n,m}}\right|\leq v_{\mathrm{max}},
\end{equation}
and that for the clearance from boundaries in Eq.~(\ref{eq:obs}):
\begin{equation}
\mathrm{dist}_{\mathcal{E}}(x_{n,m,k})\geq r.
\end{equation}
Note that satisfying the collision-free condition in Eq.~(\ref{eq:collision-free}) is a bit non-trivial. Let us define
\begin{equation}
    \mathrm{r}_q(t) := t-\left\lfloor \frac{t}{q}\right\rfloor q,
\end{equation}
and 
\begin{equation}
\begin{split}
C:=\{&(n,m,k,n',m',k')|\\
&1\leq n,n'\leq N,\; 0\leq m,m'<M,\; 0\leq k,k'<K,\\
&0 \leq \mathrm{r}_{M\tau}((m-m')\tau+k\Delta t_{n,m}-k'\Delta t_{n',m'}) < \Delta t_{n',m'},\\
&(n,m,k)\neq(n',m',k').\}.
\end{split}    
\end{equation}
Then, the collision-free condition is rewritten as, for all $(n,m,k,n',m',k')\in C$,
\begin{equation}\label{eq:dijk}
\begin{split}
d_{n,m,k,n',m',k'}:=&\left|x_{n,m,k}-((1-\alpha)x_{n',m',k'}+\alpha x_{n',m',k'+1})\right| \\ \geq& 2r,
\end{split}
\end{equation}
where
\begin{equation}
    \alpha = \frac{\mathrm{r}_{M\tau}((m-m')\tau+k\Delta t_{n,m}-k'\Delta t_{n',m'})}{\Delta t_{n',m'}}.
\end{equation}
Note that by setting $t=k\Delta t_{n,m}$, $t'=(k'+\alpha)\Delta t_{n',m'}$, the inequality in Eq.~(\ref{eq:dijk}) reduces to the original one of Eq.~(\ref{eq:collision-free}).

\subsection{Optimization}
\paragraph{Initial periodic plans.}
We first create an initial periodic plan while setting $r$ smaller than that of the original condition and $\tau$ large enough. This makes it easy to find feasible trajectories that satisfy the above conditions. Concrete algorithms used to produce such plans depend on task setups, which we present in Appendix B.

\paragraph{Objective.}
Given an initial periodic plan for $\Gamma_M= (\gamma_{n,m})_{1\leq n\leq N, 0\leq m < M}$, we optimize it with respect to $r$, $\tau$, and each trajectory $\gamma_{n, m}$ to satisfy the original conditions. We denote the original agent radius as $r_0$. By imposing the cost to violate the original conditions, solving periodic MAPP reduces to a continuous optimization problem with the following objective:
\begin{equation}
V(\tau,r,\Gamma_M):= \left(\tau-\frac{2r}{v_{\mathrm{max}}}\right)^2+ \frac{\sigma_{\mathrm{t}}}{K} \sum_{n,m,k}v_{n,m,k}^2 + c(\tau, r,\Gamma_M),
\end{equation}

\begin{equation}
\begin{split}
 c(\tau, r,&\Gamma_M):=\sigma_{\mathrm{r}}(r-r_0)^2\\
&+ \frac{\sigma_{\mathrm{v}}}{K} \sum_{n,m,k}\left(\max\left\{0,v_{n,m,k}-v_{\mathrm{max}}\right\}\right)^2\\&+
 \frac{\sigma_{\mathrm{o}}}{K}\sum_{n,m,k}\left(\max\left\{0,\mathrm{dist}_{\mathcal{E}}(x_{n,m,k})^{-1}-r^{-1}\right\}\right)^2\\
&+\frac{\sigma_{\mathrm{c}}}{K}\sum_{C} \left(\max\left\{0,d_{n,m,k,n',m',k'}^{-1}-(2r)^{-1}\right\}\right)^{2},
\end{split}
\end{equation}
where $\sigma_{\mathrm{t}}$, $\sigma_{\mathrm{r}}$, $\sigma_{\mathrm{v}}$, $\sigma_{\mathrm{o}}$ and $\sigma_{\mathrm{c}}$ are constants.
With this objective, we aim to decrease $\tau$ to the minimum $2r/v_{\mathrm{max}}$ where two agents are adjacent to each other.
We also impose costs of trajectories defined by the sums of the squares of velocity to prevent vanishing of the gradients on the trajectories.
\paragraph{Optimization method.}
We solve this optimization problem by using the Levenberg-Marquardt algorithm \cite{levenberg1944method,marquardt1963algorithm}. To force solutions to strictly satisfy the original conditions, we make the constants $\sigma_{\mathrm{r}}, \sigma_{\mathrm{v}}, \sigma_{\mathrm{p}}, \sigma_{\mathrm{c}}$ gradually increase to become large enough during the optimization. We also gradually decrease $\sigma_{\mathrm{t}}$ up to zero because the corresponding velocity term is necessary only for preventing vanishing gradients and is not included in the original conditions.

\subsection{Topological Remark} The quality of final solutions is dependent on initial periodic plans, while some initial plans will result in the same optimization results. One considerable feature of solutions is their equivalent classes with respect to continuous deformation, including optimizations, from topological perspectives.

To analyze this, we introduce an additional constraint in which no agents can pass through any start and goal locations, including those of themselves, instead of considering the conditions of the velocity and size of agents. This is necessary to ensure sufficiently different plans to be distinct enough in terms of their homotopy class. Let:
\begin{equation}
\mathcal{C}:=(\mathcal{E}\setminus\{s_1, g_1,\ldots, s_N, g_N\})\times \mathbb{R}/\mathbb{Z}.
\end{equation}
A trajectory $\gamma_{n,m}$ can be considered an embedding $\Tilde{\gamma}_{n,m}$ of an open interval $(0,1)$ to $\mathcal{C}$:
\begin{equation}
    (0,1) \ni \alpha \mapsto \left(\gamma_{n,m}(tT_{n,m}), \overline{\left(\frac{m\tau+\alpha T_{n,m}}{M\tau}\right)}\right)\in \mathcal{C}.
\end{equation}
where the overline means the equivalent class. Then, a periodic plan $(\gamma_{n,m})_{1\leq n\leq N, 0\leq m <M}$ can be considered an embedding $(\Tilde{\gamma}_{n,m})_{1\leq n\leq N, 0\leq m <M}$ of the disjoint union of $N\times M$ open intervals $(0,1)$ to $\mathcal{C}$, satisfying the following conditions for any $1\leq n \leq N$ and $0 \leq m < M$:
\begin{itemize}
    \item $\lim_{\alpha\rightarrow 0}\Tilde{\gamma}_{n,m}(\alpha)=(s_n, \overline{m/M})$ and $\lim_{\alpha\rightarrow 1}\Tilde{\gamma}_{n,m}(\alpha)=(g_n, \mathcal{A})$ for some $\mathcal{A}\in \mathbb{R}/\mathbb{Z}$.
    \item The second component of $\Tilde{\gamma}_{n,m}(\alpha)$ is locally strictly increasing with respect to $\alpha$.
\end{itemize}
Note that the collision-free condition can be interpreted as injectivity.

Now, a set of plans that are equivalent with respect to continuous deformations is the set of homotopy classes of the embeddings that satisfy the above conditions, and the following proposition holds.

\begin{proposition}
When $\mathcal{E}$ is open and connected, the above set is independent of the positions of $s_1, g_1,\ldots, s_N, g_N$.
\end{proposition}
For the proof, see Appendix A.

\section{Application to Online MAPP}\label{sec:application}
\looseness=-1
While it is assumed with periodic MAPP that the appearance timing of agents in a stream is periodic, the periodic plans once found can be used for solving a more practical problem called online MAPP, where agents appear at \emph{random times}.

\subsection{Online MAPP Problem}
While sharing certain settings with periodic MAPP, the problem of online MAPP can be viewed as a variant of online MAPF~\cite{vsvancara2019online} with the following features. Like periodic MAPP, there are several pairs of starts and goals in the environment through which streams of agents have to pass while avoiding collisions. Following online MAPF, we assume that agents can appear at random times but are also allowed to wait at their start locations in a finite or infinite queue until the subsequent agents enter the environment. This assumption is similar to the concept of `garage'~\citep{vsvancara2019online} and realistic for practical applications such as AIM.

\subsection{Proposed Method}
Adopting periodic plans to agents with random timing of appearances is straightforward. We first divide a timeline into periods of the interval $\tau$ obtained with the periodic plan. We then allow at most one agent for each period to enter each start location and follow the corresponding trajectory. Therefore, agents can avoid collisions with other agents in the same stream.

Formally, by using the notations of periodic MAPP introduced earlier, we say that an agent is assigned to the $n$-th trajectory in the $a$-th period when it waits until time $a\tau$ and moves along $\gamma_{n,a-\lfloor a/M \rfloor M}$. Let us denote as $t$, the time of appearance of a new agent at the start position $s_n$. Let $a:=\left\lceil t/\tau \right\rceil$. If the $n$-th trajectory in the $a$-th period has not yet been assigned by any agent, the new agent will follow that trajectory. Otherwise, the agent will wait to follow the next trajectory in the $a'+1$-th period, where $a'$ is the order of the period assigned to the last agent that appeared at $s_n$ before $t$ if the queue has a room. Note that if the length of the queue is finite and the number of currently waiting agents exceeds its limit, the planning for the new agent is considered a failure.

\subsection{Queueing Theoretical Analysis}
We theoretically analyze the waiting times of agents for the proposed method. 

\paragraph{Assumptions.} We assume that agents are managed by queues of infinite length. We also model time intervals of agent appearances as $c+\alpha$, where $c$ is a constant\footnote{We add $c$ to account for the time margin needed to avoid collisions between agents.} and $\alpha$ is a random variable drawn from the exponential distribution with a rate parameter $\lambda$. We also assume that all agents wait for (the maximum) time $\tau$ until they start moving. Note that this assumption corresponds to the deterministic service time in terms of the queueing theory~\cite{kendall1953stochastic} and is more conservative than in actuality.

\paragraph{Waiting time analysis.}
For theoretical analysis, we remove a constant term from the above model by subtracting $c$ from time intervals between arrivals and service time temporarily. This operation does not change the number of agents in a queue but reduces waiting times for all the agents.
The resulting model then reduces to the M/D/1 queue in terms of the queueing theory~\cite{kendall1953stochastic}. By denoting the rate of arrivals as $\lambda$ and the service time as $D:=\tau-c$, and when $\rho = D\lambda < 1$, the average waiting time is given as
\begin{equation}
W'=D+\frac{\rho}{2(1-\rho)}D.
\end{equation}
Moreover, the probability that the waiting time exceeds a limit $t$ decreases exponentially with respect to $t$~\cite{erlang1909theory}. 
By again considering $c$ we subtracted earlier, the average time now becomes
\begin{equation}
W=W'+c=\tau+\frac{\lambda (\tau-c)^2}{2(1-\lambda (\tau-c))}.
\end{equation}

\section{Experiments}\label{experiments}

\begin{figure*}[t]
    \centering
    \includegraphics[width=0.117\textwidth]{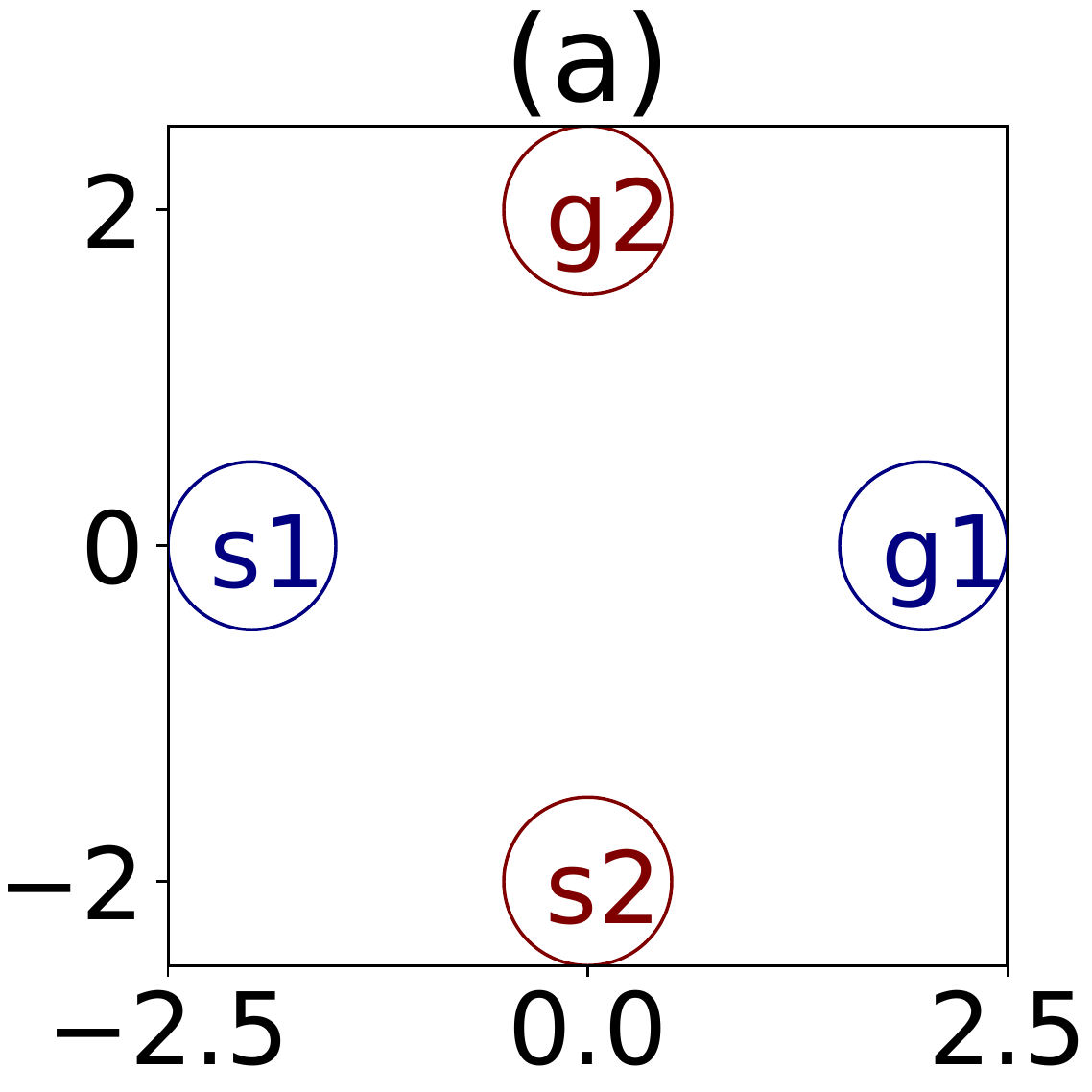}
    \includegraphics[width=0.135\textwidth]{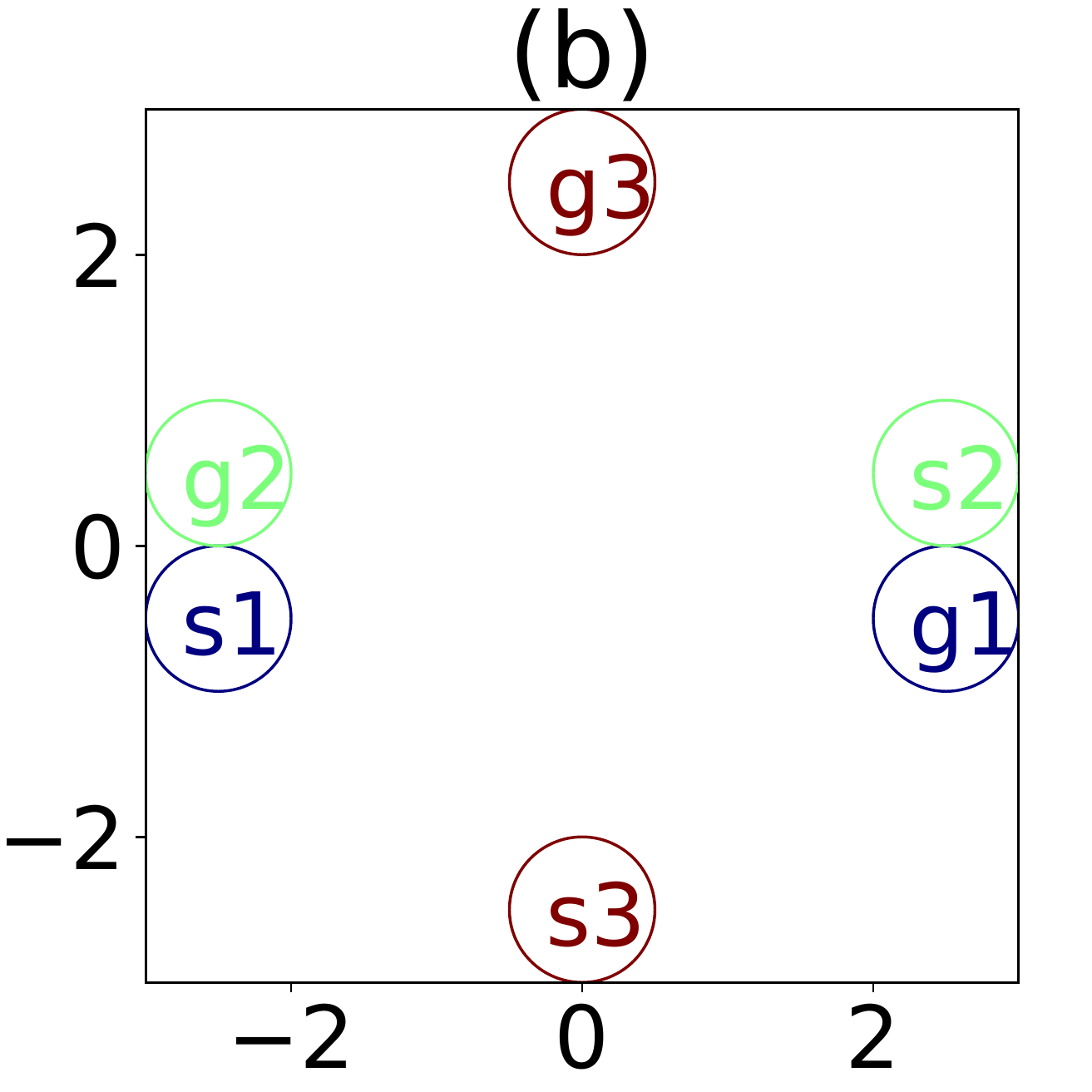}
    \includegraphics[width=0.135\textwidth]{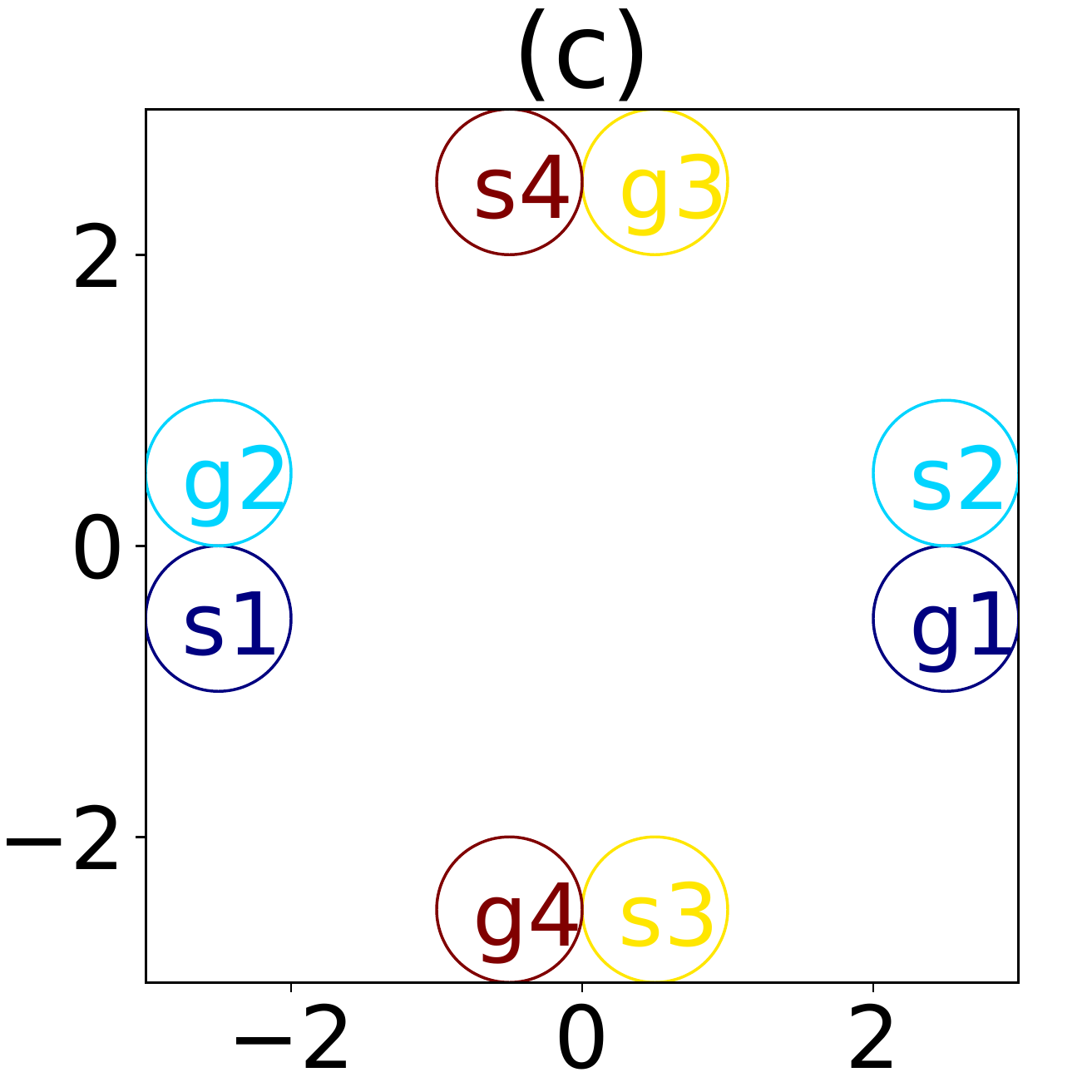}
    \includegraphics[width=0.171\textwidth]{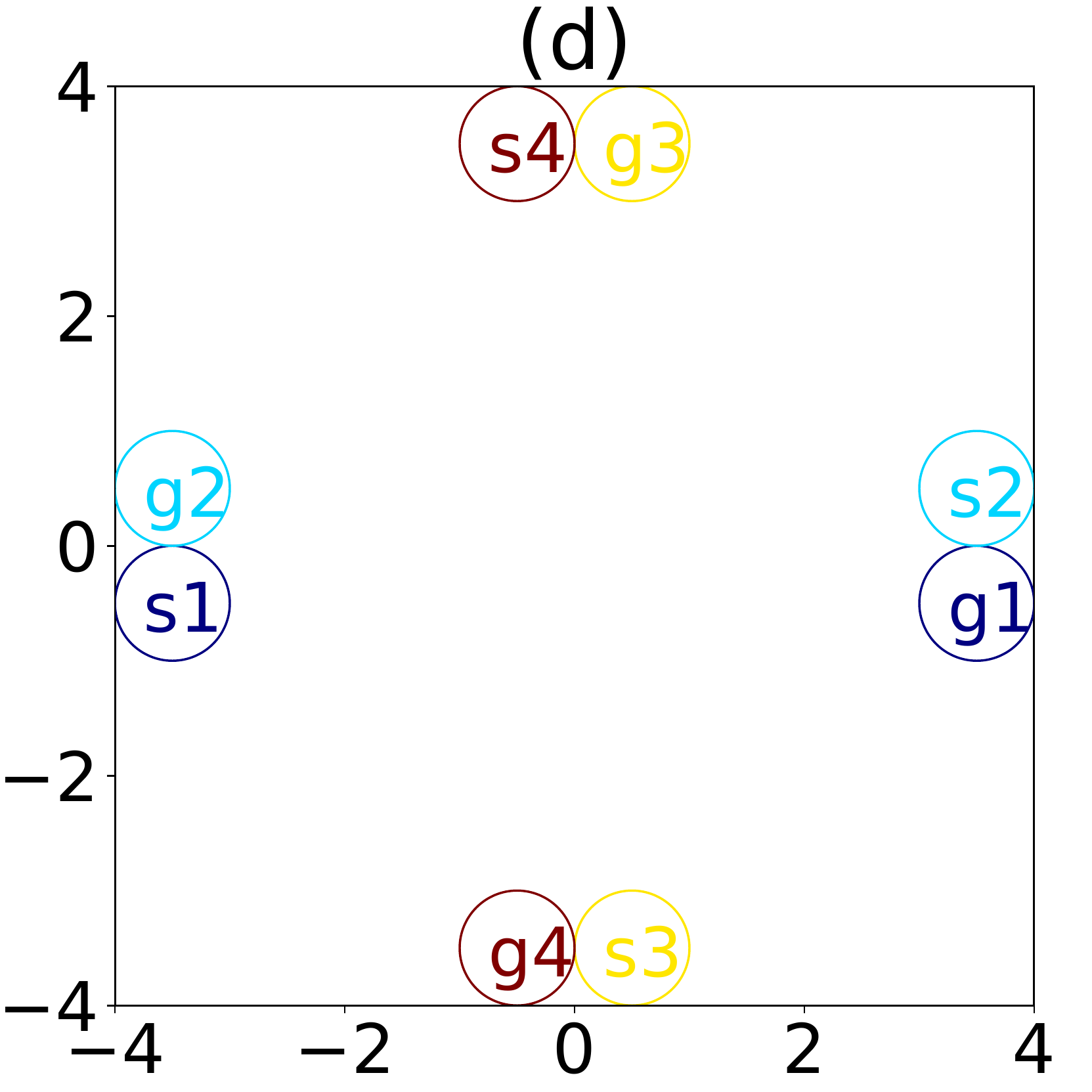}
    \includegraphics[width=0.171\textwidth]{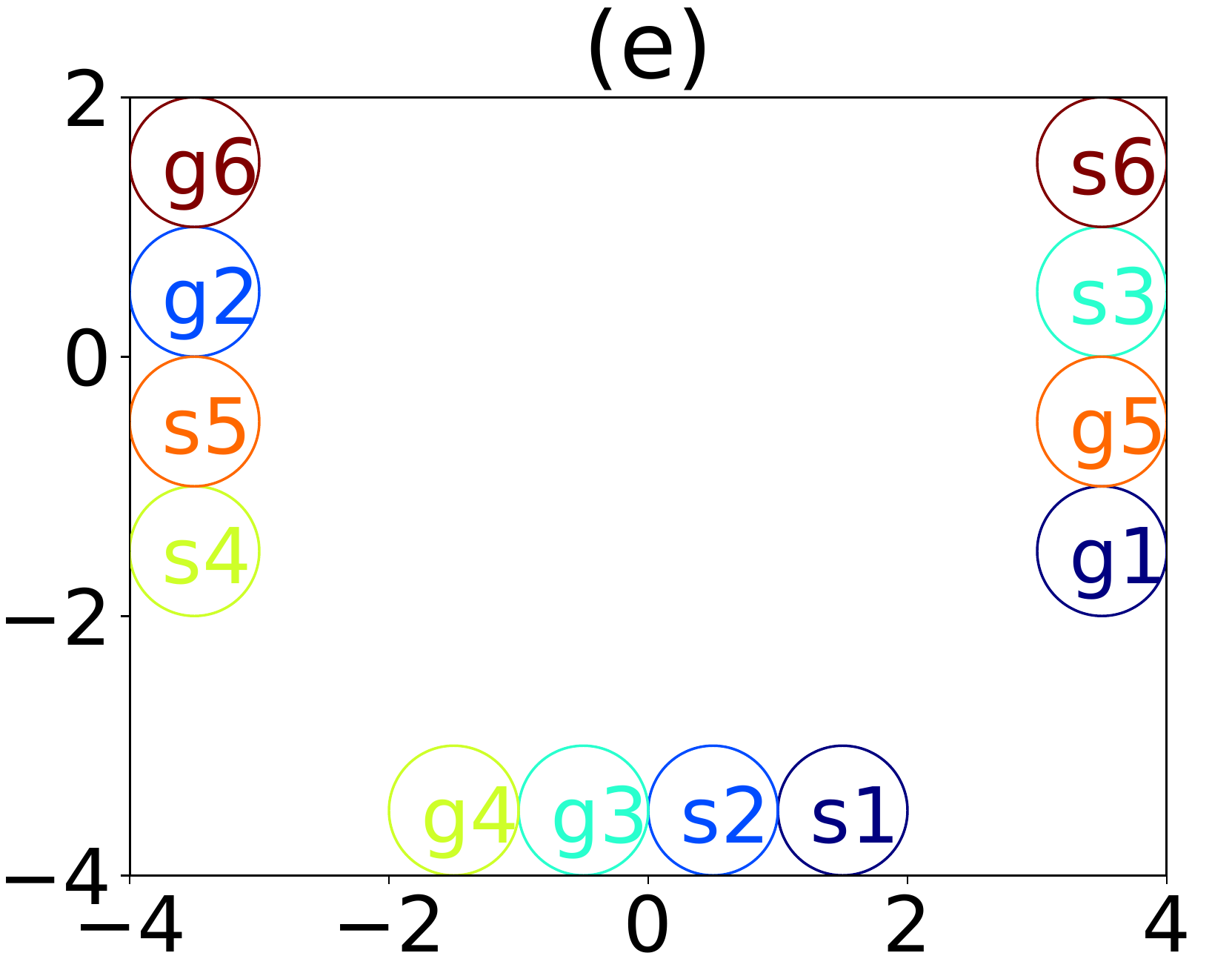}
    \includegraphics[width=0.207\textwidth]{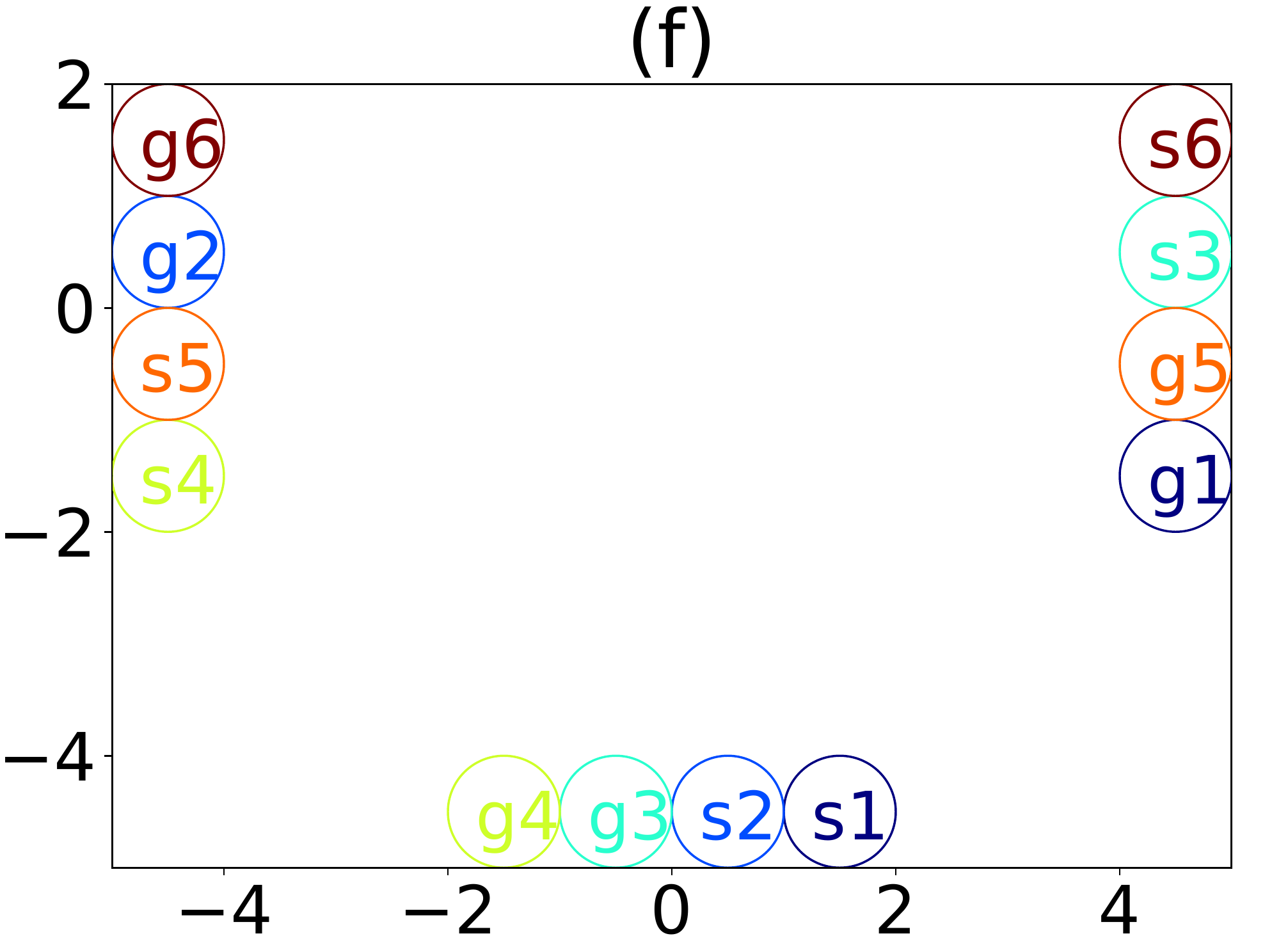}
    \caption{Environments that abstract AIM tasks. Letters `s' and `g' indicate start and goal locations, respectively.}
    \label{fig:environments}
\end{figure*}

\begin{figure*}[t]
    \centering
    \includegraphics[width=0.117\textwidth]{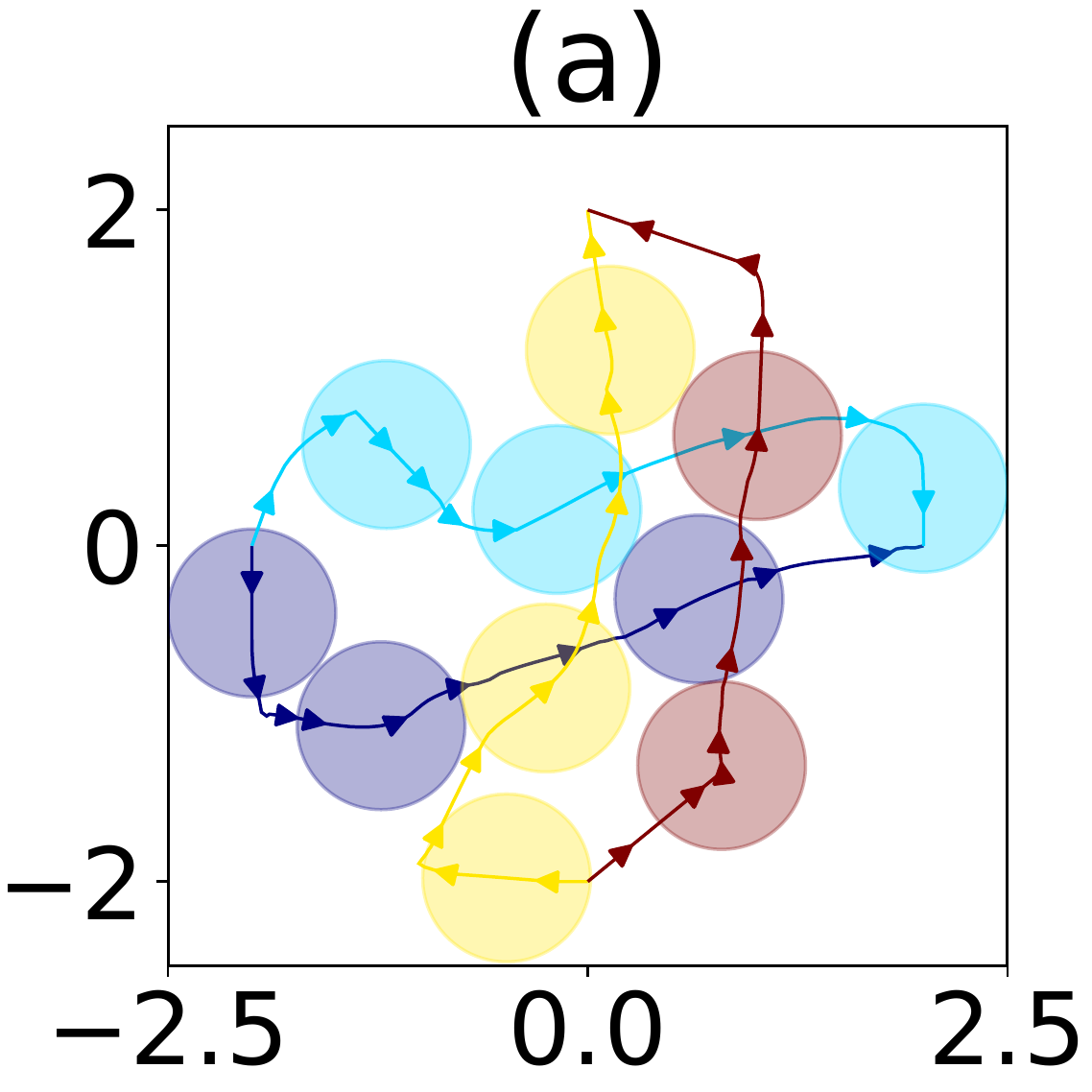}
    \includegraphics[width=0.135\textwidth]{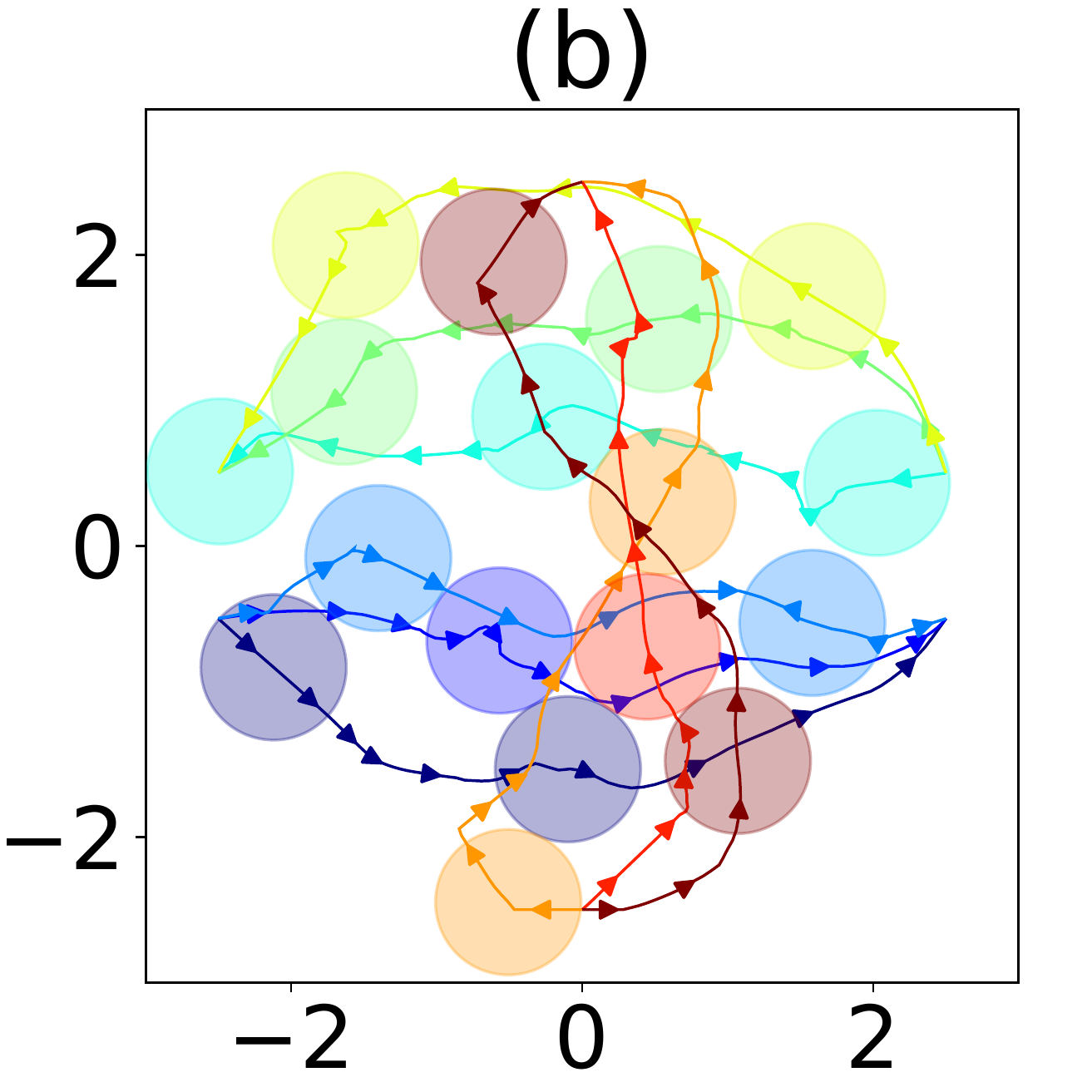}
    \includegraphics[width=0.135\textwidth]{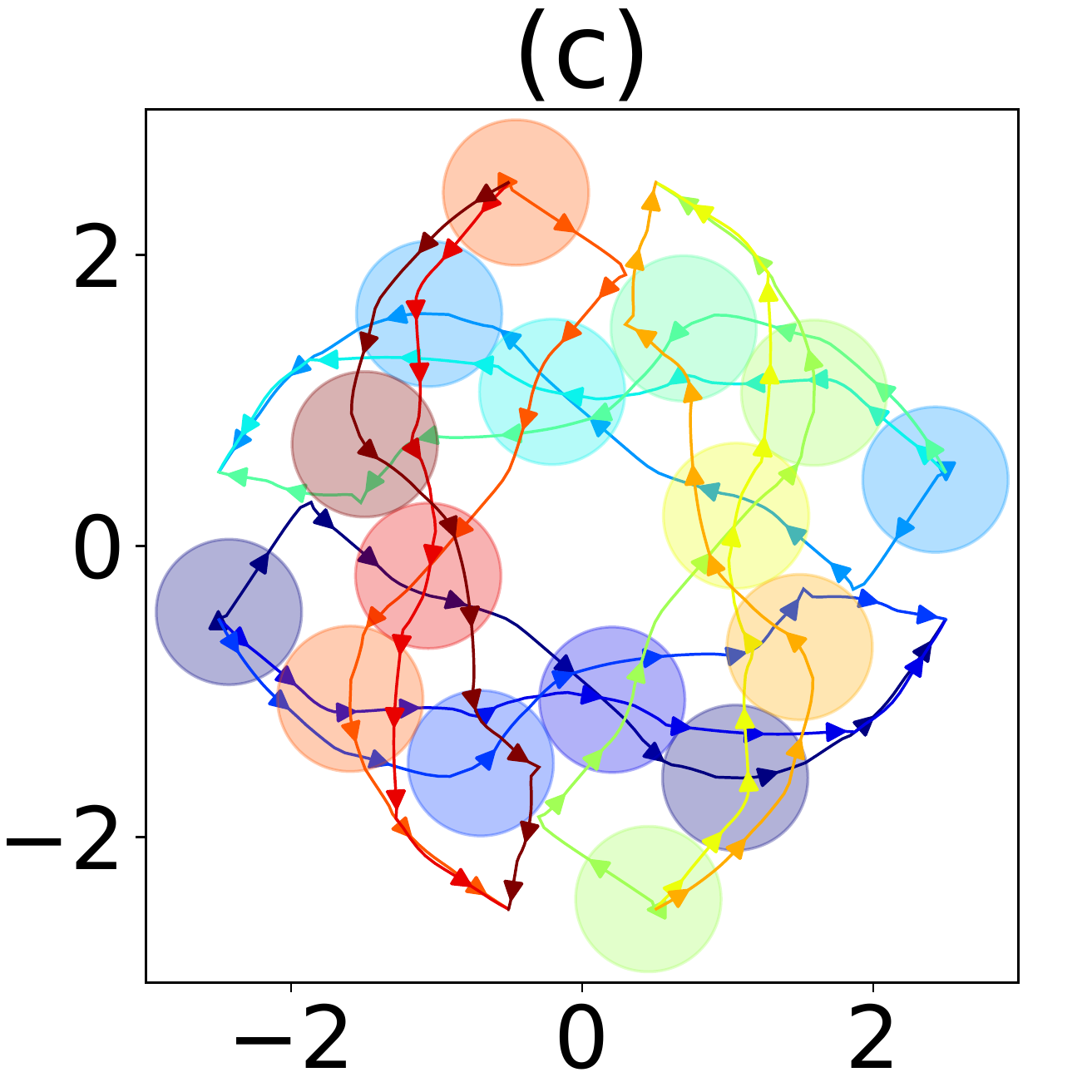}
    \includegraphics[width=0.171\textwidth]{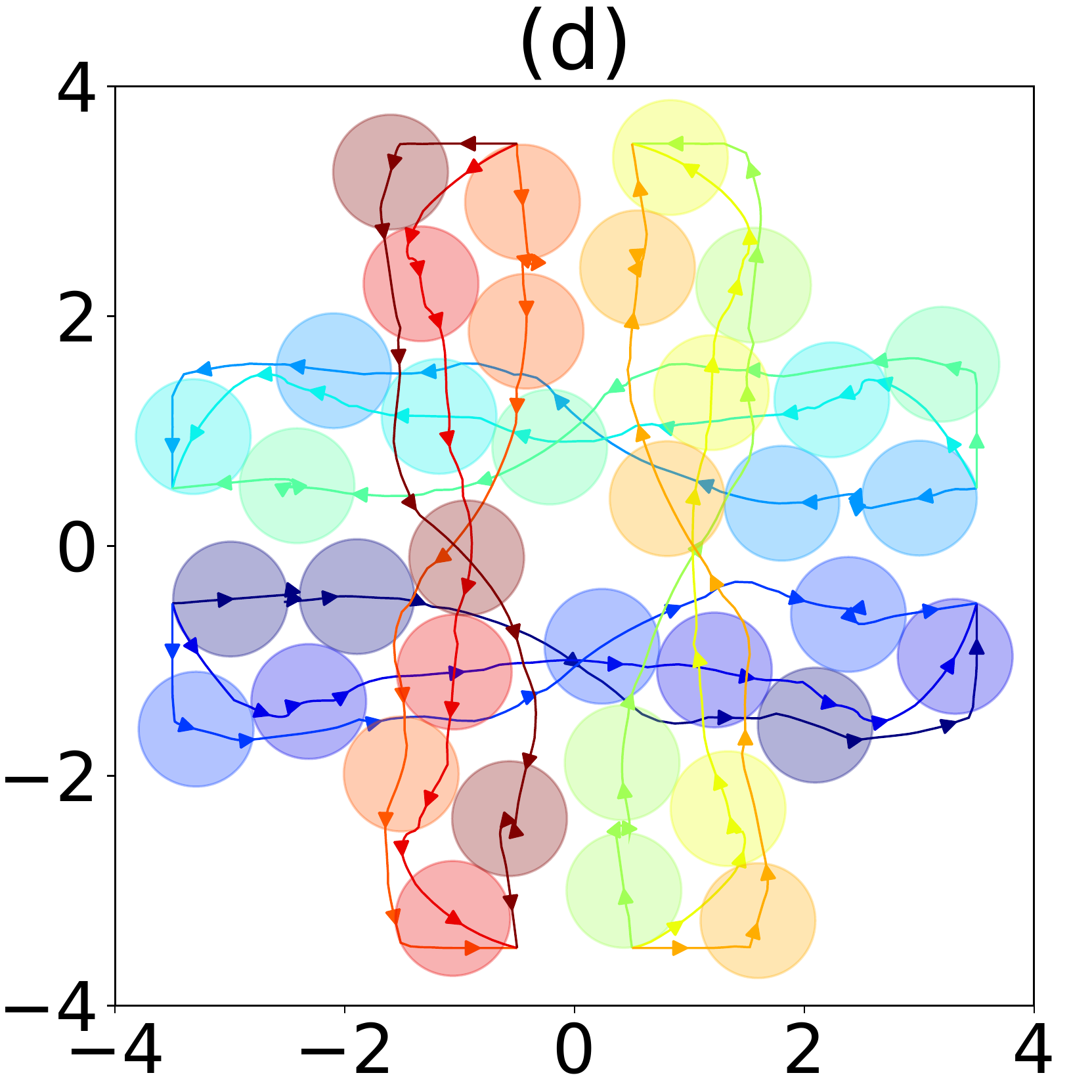}
    \includegraphics[width=0.171\textwidth]{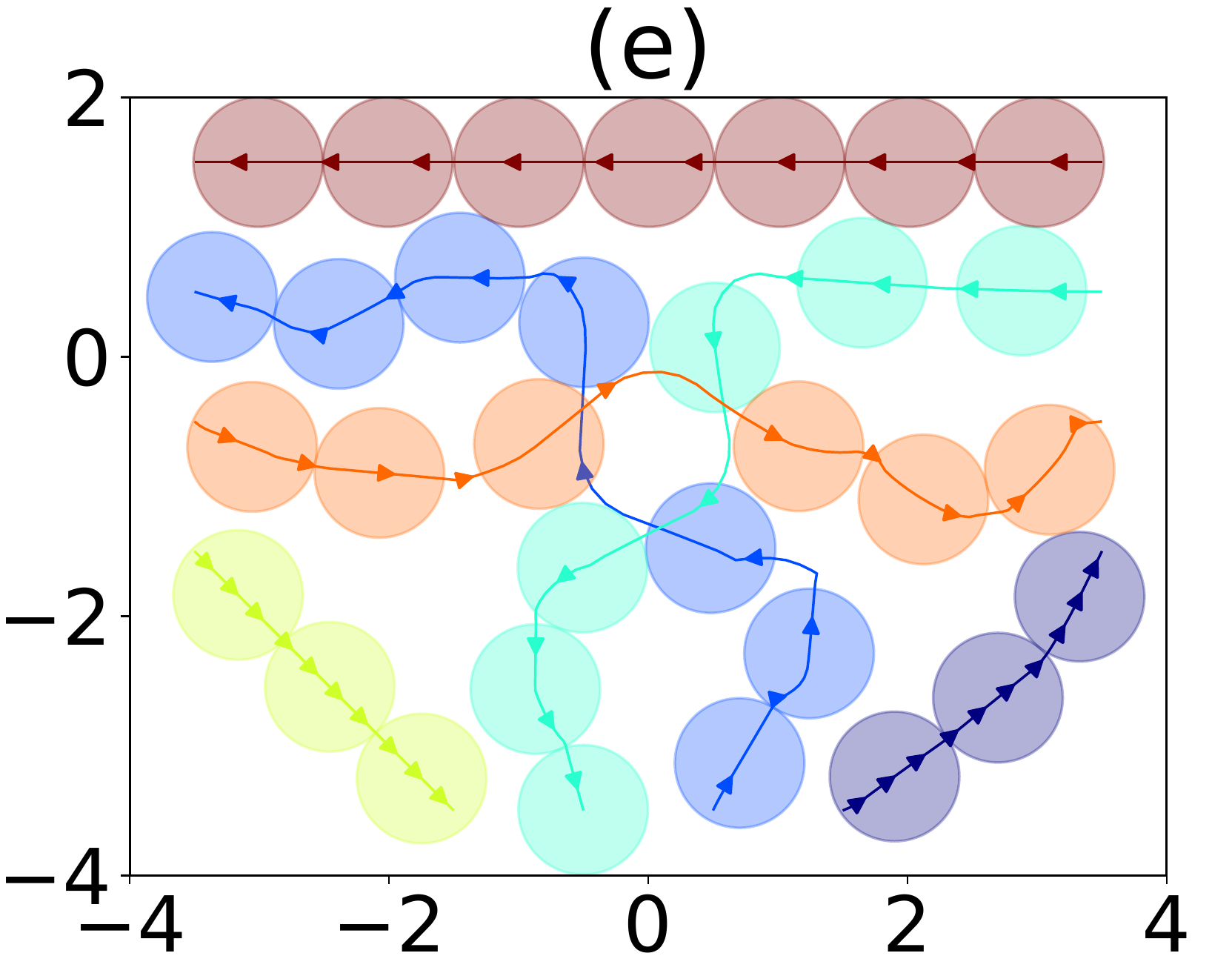}
    \includegraphics[width=0.207\textwidth]{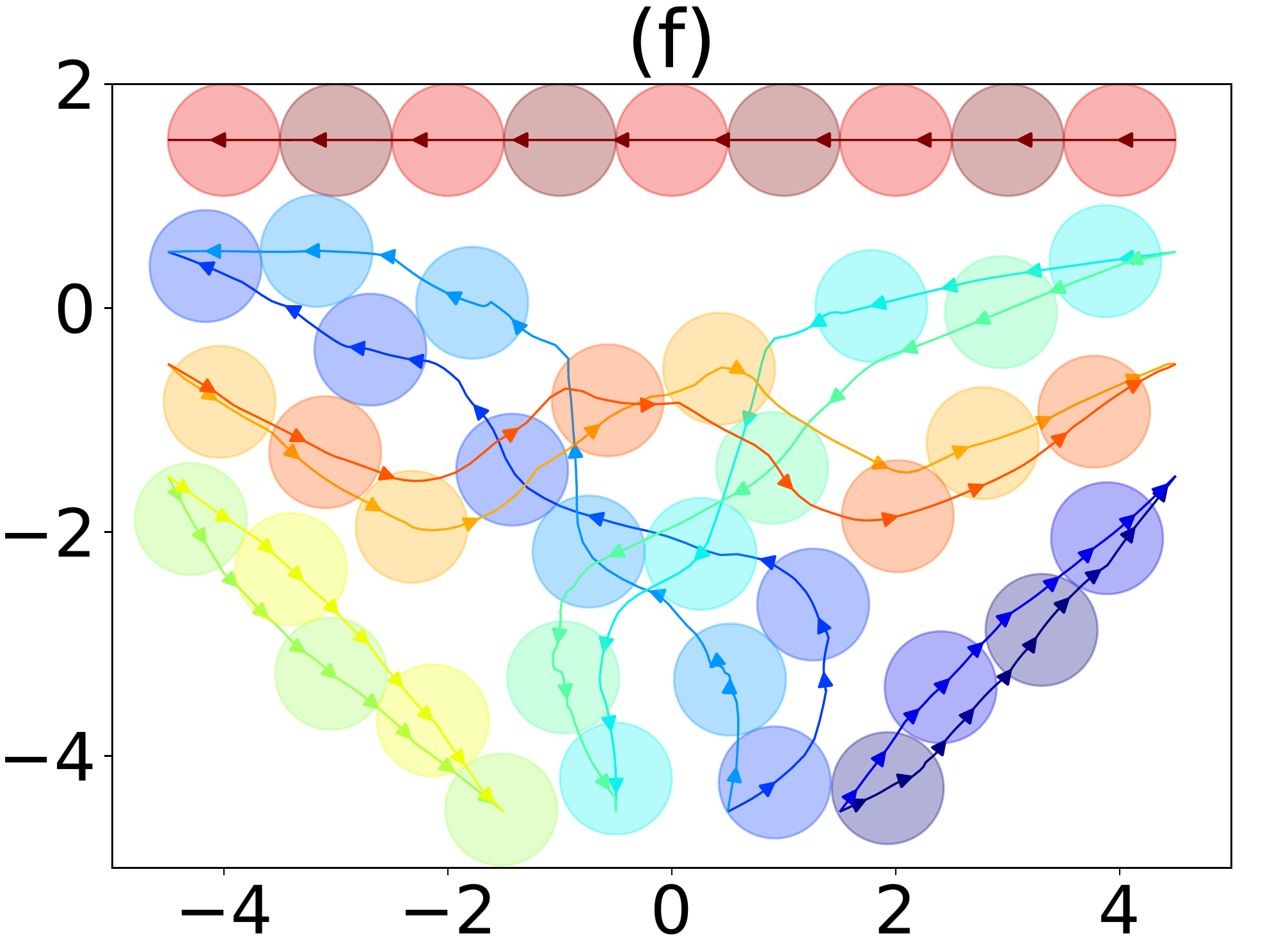}
    \caption{Optimized periodic plans for each environment with best $M$. Best viewed in videos in the Appendix.}
    \label{fig:best periodic plans}
\end{figure*}

\begin{figure}[t]
    \centering
    \includegraphics[width=0.5\linewidth]{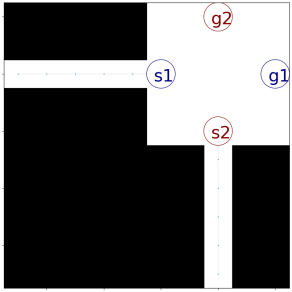}
    \caption{Environment (a) with corridors used as queues for five agents.}
    \label{fig:corridors}
\end{figure}

To evaluate the effectiveness of our method for first solving periodic MAPP then using periodic plans for online MAPP problems, we focus on scenarios of abstracting AIM tasks~\cite{dresner2008multiagent}.

\subsection{Experimental Setups}
\paragraph{Environments.}
Our AIM scenario involves a single intersection with several entrances and exits. Figure~\ref{fig:environments} shows the six environments used in the experiments. Each environment abstracts one of the typical situations of intersections with different sizes. The circles with letters `s' and `g' are the start and goal locations, respectively. Specifically, environment (a) is a crossing of two one-way roads, while environment (b) is a crossing of one-way and two-way roads. Environments (c) and (d) are modeling the crossing of two two-way roads of different sizes. Finally, environments (e) and (f) model the T-junction with different sizes. We also assume that each start location is equipped with a `corridor', as illustrated in Fig.~\ref{fig:corridors}, which we can use as a queue (with unlimited or limited capacity) to keep agents waiting until entering the intersection. Agents (\ie, vehicles) are modeled by a circle and follow a simple holonomic kinodynamics model that enable them to move in any direction under a given maximum velocity. Trajectories for a new agent must be planned immediately once that agent appears in the environment. Nevertheless, it is possible to replan trajectories for some agents that have already been moving in the environment to take into account the new agent.

\paragraph{Parameters.}
We evaluated two different configurations for the queues: unlimited or limited capacity with five agents at most. The time interval between agent appearances is modeled as $1.0 + \alpha$, where $\alpha$ follows the exponential distribution with a rate parameter $\lambda$. We sampled appearance times until the last time reached $1000$ for infinite queues and $100$ for finite queues. For each environment and each $\lambda\in [0.25, 0.5, 0.75,\dots,2.5]$, we generated $10$ different problem instances. Throughout the experiment, $r$ and $v_{\mathrm{max}}$ were respectively fixed to $0.5$ and $1.0$.

\paragraph{Evaluation metrics.}
The quality of plans is measured by the following two metrics.
\begin{itemize}
\item \textbf{Throughput} measured as the number of agents entering the environment in unit time.
\item \textbf{Average delay} calculated as the incremental travel time compared with the shortest possible trajectory averaged over agents. Note that this metric includes the time for the agent waiting in a queue.
\end{itemize}

\subsection{Initial Periodic Plans}
With periodic MAPP, we generated and evaluated three periodic plans for each problem instance with cycles $M \in \{1, 2, 3\}$.
We constructed the plans by carefully designing the order of passing at intersection points such that, after $M$ agents of one direction pass at the intersection, $M$ agents of another direction pass alternately. 

Formally, the initial plan for $M$ was created on the basis of the following rules:
\begin{itemize}
    \item Let $\pi_n$ be the shortest path connecting $s_n$ and $g_n$. Each agent appearing at $s_n$ follows $\pi_n$, while adjusting its velocity to satisfy the next condition. 
    \item Let two paths $\pi_n$ and $\pi_{n'}$ intersect at point $p$. We assume that the length of the part of $\pi_n$ from $s_n$ to $p$ is shorter than that of the part of $\pi_{n'}$ from $s_{n'}$ to $p$ (and assume $i<i'$ if they are equal). Then, for any $a \in \mathbb{Z}$, $M$ agents appearing at $s_n$ at time $aM\tau, (aM+1)\tau, \ldots, (aM+M-1)\tau$ must pass $p$ at the time between times when agents appearing at $s_{n'}$ at $(aM-1)\tau$ and at $aM\tau$ pass $p$. The differences between times when two agents pass are set to be not smaller than $1.0$ to prevent optimization failures.

\end{itemize}
Note that such initial plans can always be constructed as long as the initial $\tau$ is taken to be large and $r$ is small enough. See Appendix B for details of the generation algorithm.

\subsection{Baseline Methods}

Because the same problem setup (\ie, improving throughput of agent streams for MAPP in continuous space and time) has not been explicitly addressed, we developed two baseline methods called \emph{first-come and first-serve (FCFS)} and \emph{snapshot optimal (SO)}, which combine general strategies of online MAPF~\cite{vsvancara2019online} and the state-of-the-art MAPP algorithms in continuous space and time~\cite{andreychuk2022multi,kasaura2022prioritized}. For both baselines, we first used probabilistic roadmap with the fixed number of neighbors (k-PRM)~\cite{karaman2011sampling} as a standard approach to approximate the continuous space into a roadmap and find collision-free paths on the constructed roadmap.
Implementation details are presented in Appendix C.
\begin{itemize}
    \item \textbf{FCFS}: This baseline incrementally plans a collision-free trajectory for every new agent that appears while regarding trajectories of other agents already present in the environment as space-time obstacles. This is a natural application of prioritized planning~\cite{silver2005cooperative} for the ``replan single'' strategy introduced in the context of online MAPF~\cite{vsvancara2019online}. Specifically, we used prioritized safe-interval path planning~\cite{kasaura2022prioritized} to handle continuous space and time.
    \item \textbf{SO}: By contrast, this baseline uses the ``replan all'' strategy~\cite{vsvancara2019online} and finds the optimal solution using continuous conflict-based search~\cite{andreychuk2022multi} for all the agents presenting in the environment each time a new agent appears.
\end{itemize}

\subsection{Results}

\paragraph{Optimization results.}
Table~\ref{tab:periods} shows the periods of optimized plans for cycle $M=1, 2, 3$ and Fig.~\ref{fig:best periodic plans} illustrates the optimized periodic plans with the best choice of $M$. We would like to emphasize that this optimization is required only once for each combination of environments and $M$. Unlike the other baselines, our method can be used without requiring agents to communicate with other agents to retrieve their current locations or replanning to adjust their trajectories for each appearance of new agents.

\begin{table}[t]
    \centering
    \begin{tabularx}{0.85\linewidth}{l|cccccc}
    \toprule[1pt]
    & \multicolumn{6}{c}{\textbf{Environment}}\\
    $M$&(a)&(b)&(c)&(d)&(e)&(f)\\
    \midrule
         $1$& 2.03& 2.08& 2.45& 2.44 &2.34& 2.13\\
    \midrule
         $2$& 1.27& 1.54& 1.87& 1.86 &3.02& 1.86\\
    \midrule
         $3$& 1.56& 1.45& 1.86& 1.63 &3.04& 3.54\\
    \bottomrule[1pt]
    \end{tabularx}
    % \caption{Periods of optimized plans.}
    \caption{Periods for optimized periodic plans with different cycle $M$}
    \label{tab:periods}
\end{table}

\begin{figure}[t]
    \centering
    \includegraphics[width = \linewidth]{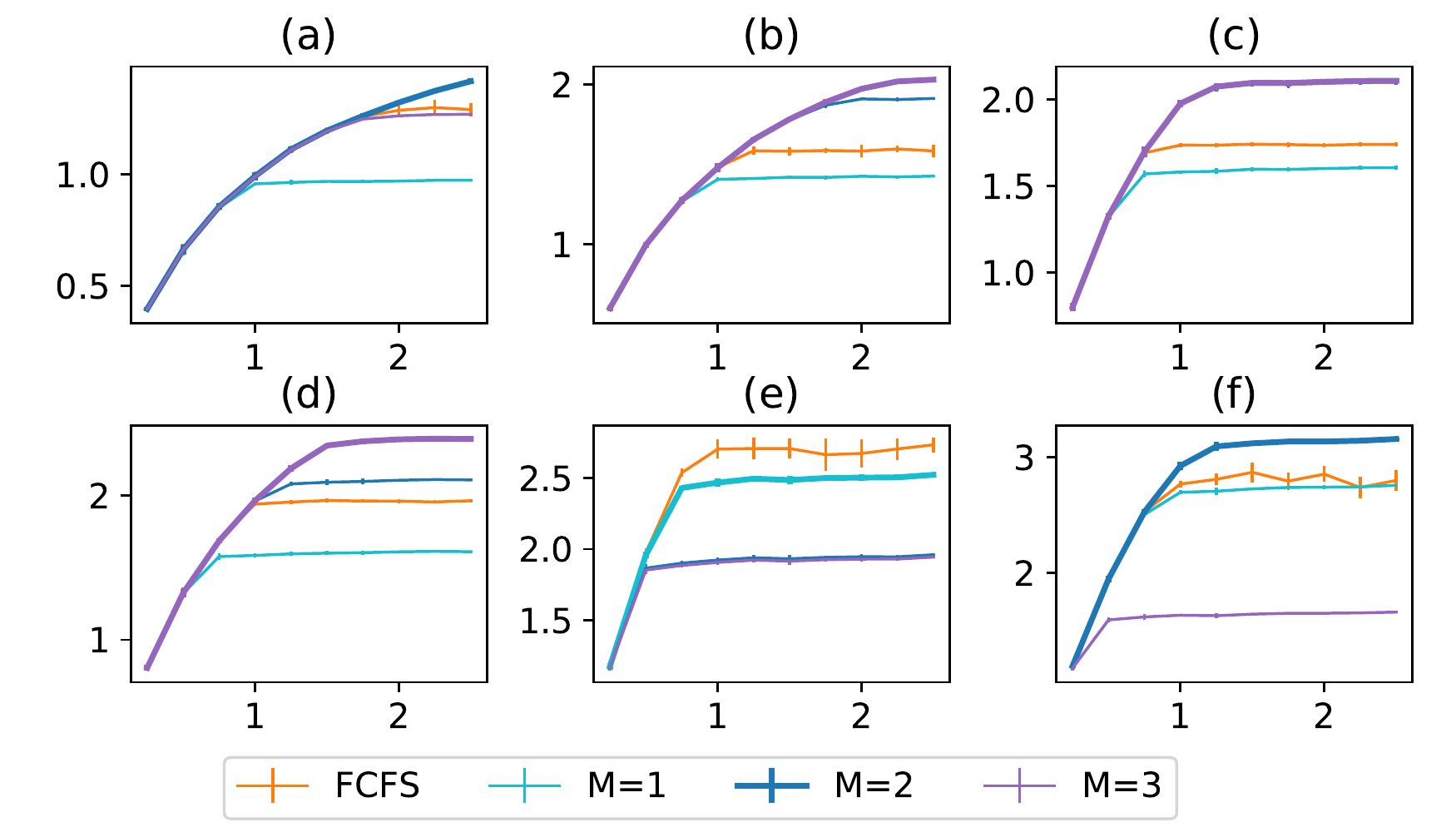}
    \caption{Throughput with respect to $\lambda$ for \emph{infinite} queues (bars represent standard derivations)}
    \label{fig:throughputs}
\end{figure}

\begin{figure}[t]
    \centering
    \includegraphics[width = \linewidth]{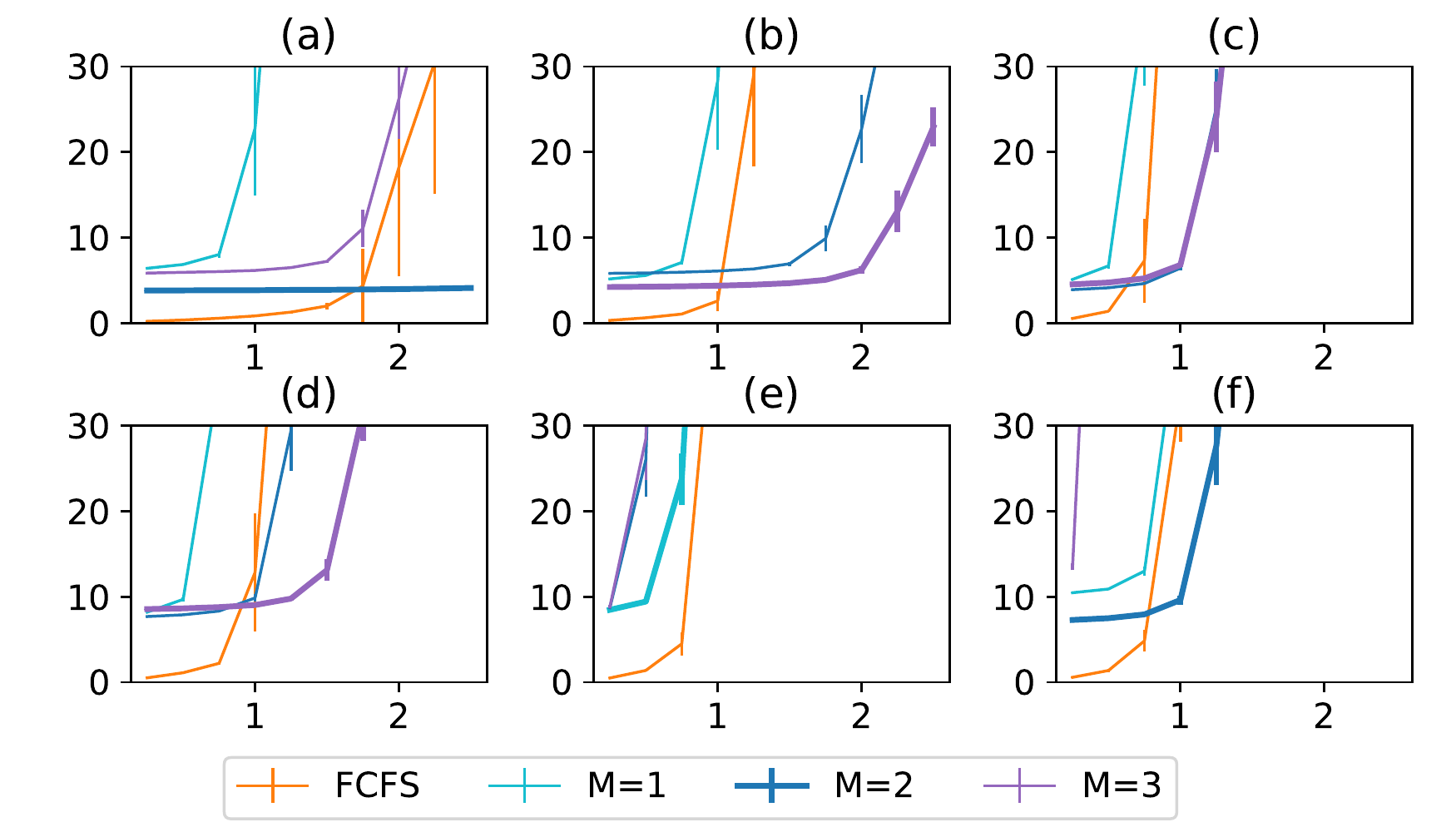}
    \caption{Average delay with respect to $\lambda$ for \emph{infinite} queues (bars represent standard derivations)}
    \label{fig:delay 2}
\end{figure}

\paragraph{Case I: Infinite queues.}
Figures~\ref{fig:throughputs} and \ref{fig:delay 2} respectively show the changes in throughput and average delays with respect to $\lambda$ when the capacity of queues is infinite. With the exception of the environment (e), our method, especially for $M=2$ or $M=3$, consistently outperformed FCFS as $\lambda$ became higher. Note that it was not possible to use SO because the number of agents present in the environment could quickly become very large. We observed that the limited performance of the proposed method for (e) is possibly due to the difficulty of finding enough room for periodic intersections in the environment. 

\begin{figure}[t]
    \centering
    \includegraphics[width = \linewidth]{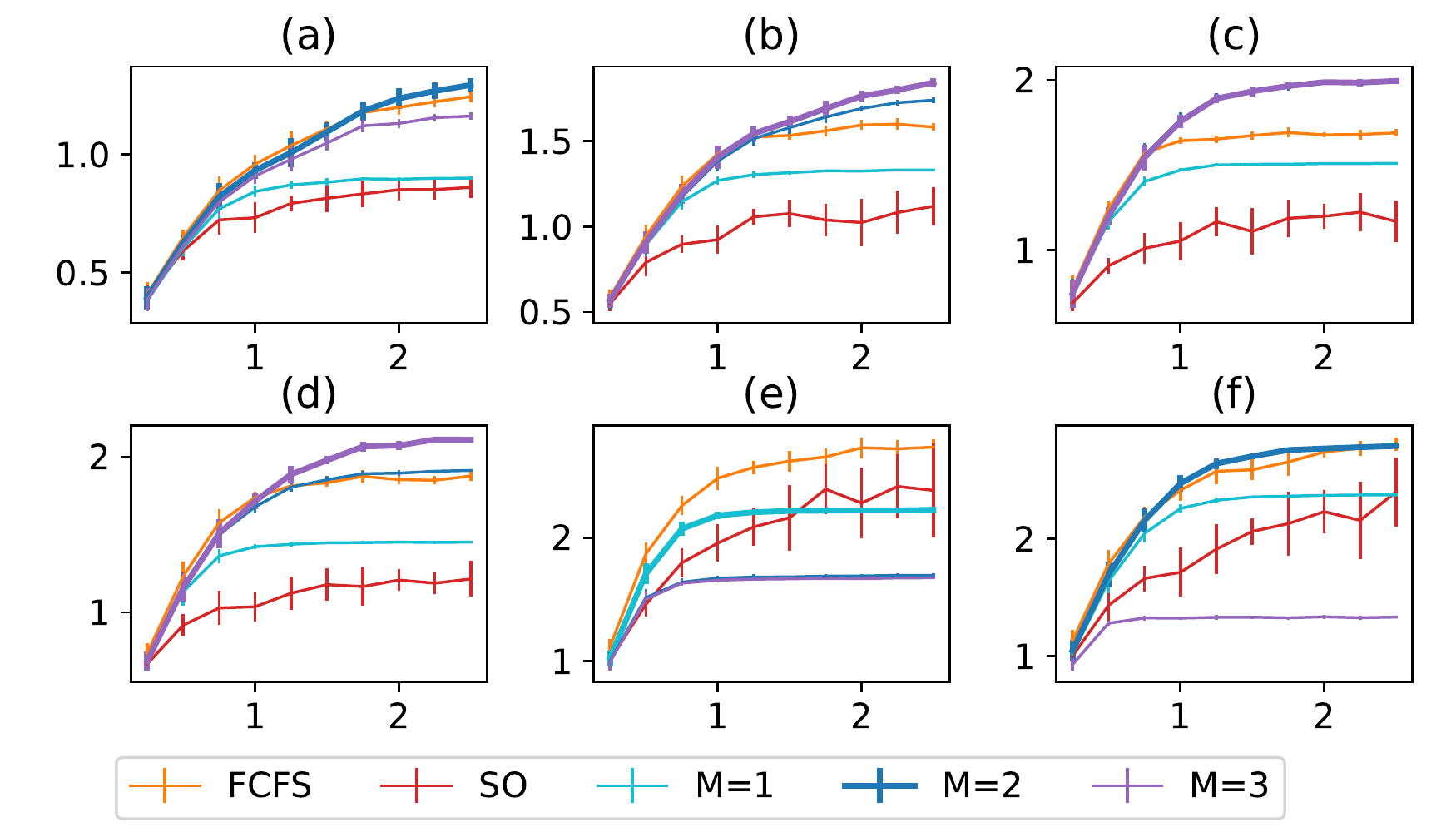}
    \caption{Throughput with respect to $\lambda$ for \emph{finite} queues (bars represent standard derivations)}
    \label{fig:success rates}
\end{figure}

\begin{figure}[t]
    \centering
    \includegraphics[width = \linewidth]{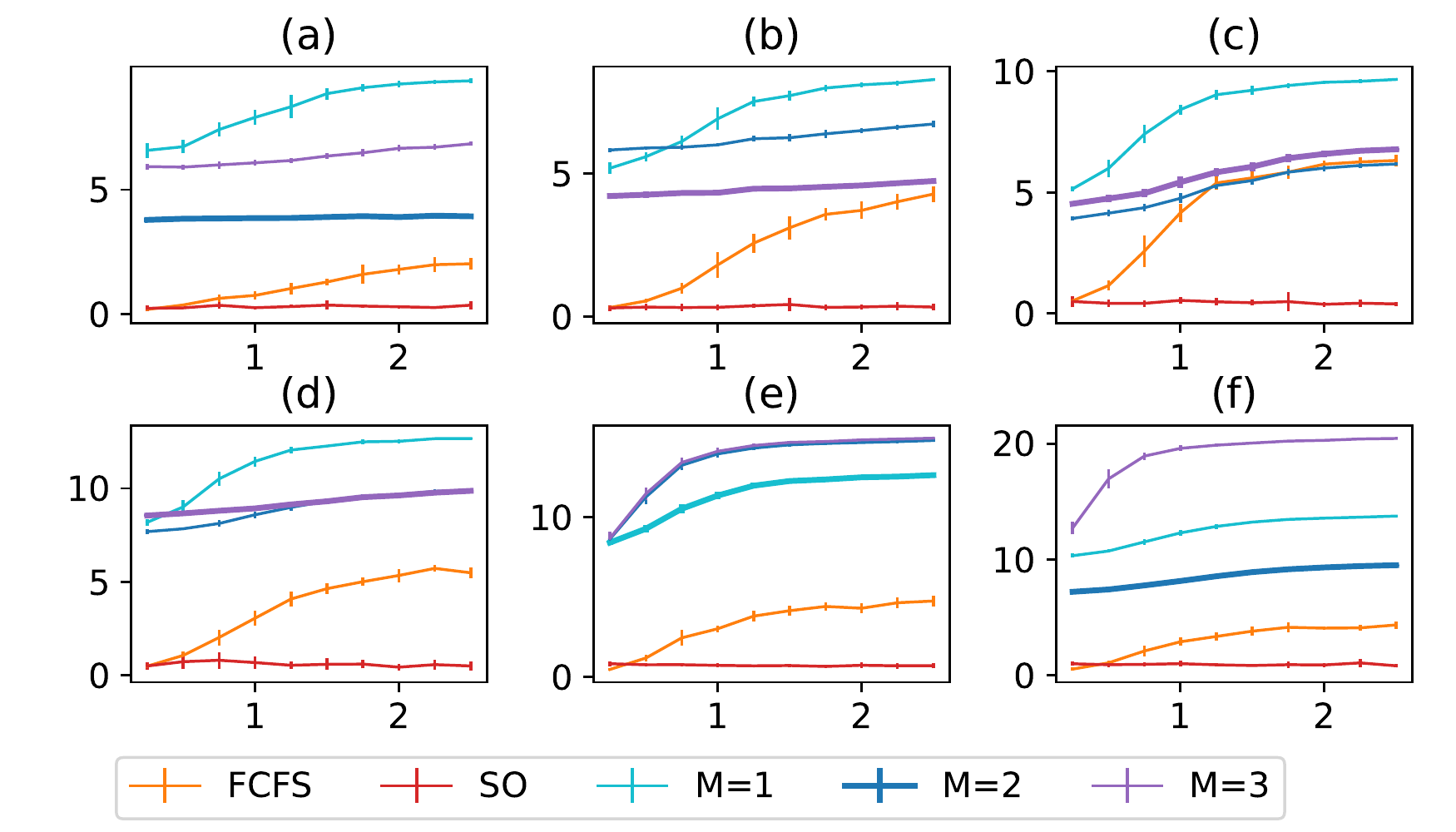}
    \caption{Average delay with respect to $\lambda$ for \emph{finite} queues, (bars represent standard derivations)}
    \label{fig:delay 1}
\end{figure}

\paragraph{Case II: Finite queues.}
We set the length of queues to be five. Note that planning for a new agent is regarded as failed if the number of agents waiting to enter their starts exceeded the length of queues or when the planning time exceeds a predefined limit (1.0 s). Agents resulting in such planning failures are discarded and degrade throughput. Figures~\ref{fig:success rates} and \ref{fig:delay 1} respectively show the throughput and average delays changing with respect to $\lambda$. Regarding throughput, our method for $M=2$ and $M=3$ performed better than FCFS and clearly outperformed SO, except for environment (e). The performance of SO degraded mainly due to the planning failures. By contrast, our method showed limited performances in terms of average delays, indicating the limitation of our method that trajectories obtained using periodic plans would be relatively redundant compared with those from FCFS and SO.

\section{Related Work}

\paragraph{MAPP and its variants.} 
MAPP has historically been studied in the fields of artificial intelligence and robotics. Many studies have considered planning particularly in discrete spaces such as grid maps, which are often referred to as MAPF~(\eg, \citet{silver2005cooperative,sharon2015conflict}, and see \citet{stern2019multi} for an extensive survey.) There are a number of variants of MAPF problems, such as Anonymous MAPF (\ie, no correspondences between starts and goals)~\cite{Kloder2006,Yu2013}, MAPF with kinodynamic constraints~\cite{honig2018trajectory,ijcai2018-74}, and online MAPF~\cite{vsvancara2019online,ma2017lifelong,Li2021}, also known as lifelong MAPF. As summarized by \citet{Ma2021}, online MAPF is an online version of MAPF in which a team of agents is asked to solve a stream of tasks. Agents are assigned a new task whenever they appear in the environment~\cite{vsvancara2019online} or upon reaching their goal~\cite{ma2017lifelong}. In contrast, periodic MAPP and our version of online MAPP are different in that streams of agents are asked to solve a certain task, with the unique objective that aims for high throughput. We introduced a challenging problem setup that solves online MAPP in the continuous space and with continuous time. Studies on such continuous setups have significantly been limited compared with discrete cases~\cite{walker2018extended,honig2018trajectory,andreychuk2021improving,andreychuk2022multi,kasaura2022prioritized,okumura2022ctrms}.

\paragraph{Application to AIM.} As reviewed by \citet{stern2019multi,Ma2021}, AIM is a common application of online MAPF. Typical approaches include the first-come and first-serve for repeatedly determining trajectories for every new agent~\cite{dresner2008multiagent} and the application of optimal solvers for a set of all agents present at the moment~\cite{vsvancara2019online}, which we compared in our experiments. For a given trajectory (or simply lane), there are studies that used deep reinforcement learning to achieve an optimal policy for vehicle acceleration control~\cite{Kreidieh2018,Jang2019,Cui2021}.

\paragraph{Optimization for MAPP.}
\looseness=-1
Similar to this work, other studies uses continuous optimization of trajectories (\ie, trajectory deformation~\cite{kurniawati2007path}), especially for single-agent cases. For example, the idea of avoiding collisions on the basis of the optimization has existed for a long time~\cite{khatib1986real}. Another widely used motivation for optimization is to take into account kinodynamic constraints, \eg, that in \citet{rosmann2017kinodynamic} for a single-agent case and that in \citet{honig2018trajectory} for a multi-agent case. Our study is unique in that we optimized trajectories as well as the period for their repeated use.

\paragraph{Other related work.}
Finally, our study has several more connections to other prior works. Solving path planning while relaxing constraints is a technique for single-agent cases~\cite{bonilla2015sample,bonilla2017noninteracting,fusco2018improving}.
The homotopical aspect of path planning has been studied by \citet{bhattacharya2010search} for single-agent and by \citet{bhattacharya2018path} for multi-agent cases. However, these studies are not directly applicable to our problem setup due to the difficulty of considering collisions for agents appearing in a periodic fashion.

\section{Conclusion}
We presented a new variant of MAPP called periodic MAPP in which a stream of agents can appear periodically at each start location and leave the environment once they arrive at the goal. We also proposed a solution method of periodic MAPP that generates a periodic plan, \ie, a set of collision-free trajectories that can be used repeatedly over periods while maintaining high throughput. We showed that the periodic plans can further be used for solving the online MAPP problem in which agents in each stream appear at a random time, and demonstrated its effectiveness on scenarios of abstracting AIM tasks.

% \paragraph{Future directions.}
Currently, our formulation of periodic MAPP as well as our solution method can cover only situations in which agents follow a simple kinodynamics model that takes into account only maximum velocity. Promising future work is to address more realistic kinodynamics of wheeled robots or drones. This would require extending an optimization technique for solution methods such as ones used to plan trajectories for swarms~\cite{honig2018trajectory}. Another interesting direction from an application perspective is to tackle a more challenging AIM task in which human-driven vehicles also exist. In such a case, it will be important to combine the proposed method with state-of-the-art AIM methods that can reactively control acceleration on a given trajectory~\cite{Kreidieh2018,Jang2019,Cui2021}.

%\balance
\clearpage

%\balance
\clearpage

\section{A: Proof of Proposition 1}
Let $s_1', g_1',\ldots, s_N', g_N'$ be another collection of start and goal locations. %We can assume that $s_1, g_1, \ldots, s_N, g_N, s_1', g_1',\ldots, s_N', g_N'$ are all distinct. Let $S:=\{s_1, g_1,\ldots, s_N, g_N, s_1', g_1',\ldots, s_N', g_N'\}$.
It is enough to show that there exists a homeomorphism $F:\mathcal{E} \rightarrow \mathcal{E}$ such that $F(s_n)=s_n'$ and $F(g_n)=g_n'$ for all $1\leq n \leq N$. This is proven by \citet{michor1994n}.

\section{B: Generation Algorithm of Initial Plans}
First, we set $r=0$ to relax the clearance from boundaries and collision-free constraints.
Let $s_n=p_{n, 0}, p_{n, 1}, \ldots, p_{n, K_n}=g_n$ be start and goal locations and all intersection points with other paths on $\pi_n$ ordered from $s_n$ to $g_n$. Also, let $t_{n,m,k}$ be the traveling time from $s_n$ to $p_{n,k}$ in the initial plan, and $L_{n,k}$ be the minimum traveling time from $p_{n,k}$ to $p_{n,k+1}$.
Then, the following condition must hold:
\begin{equation}\label{eq:zero condition}
    t_{n,m,0}=0,
\end{equation}
\begin{equation}\label{eq:L condition}
    t_{n,m,k}+L_{n,k}\leq t_{n,m,k+1},
\end{equation}

We also consider constraints of the forms ``the $m$-th agent appeared $s_n$ must pass through $p_{n,k}=p_{n',k'}$ earlier than $m'$-th agent appeared $s_n$'', where $0\leq m ,m'< M$, which is written by:
\begin{equation}
    m\tau+t_{n,m,k}< m'\tau+t_{n',m',k'}.
\end{equation}
We describe these constraints using tuple $C_1:=(n,m,k,n',m',k')$.
Likewise, there exist constraints of the forms ``the $(M-1)$-th agent appeared $s_n$ must pass through $p_{n,k}=p_{n',k'}$ earlier than $M$-th agent appeared $s_{n'}$''. They are given as follows:
\begin{equation}
    (M-1)\tau+t_{n,M-1,k}< M\tau+t_{n',0,k'},
\end{equation}
which we describe them using tuple $C_2:=(n,k,n',k')$.

Moreover, we set a redundancy parameter $R=1.0$ and make these constraints stricter:
\begin{equation}\label{eq:C1 condition}
    m\tau+t_{n,m,k}+R\leq m'\tau+t_{n',m',k'}.
\end{equation}
\begin{equation}\label{eq:C2 condition}
    (M-1)\tau+t_{n,M-1,k}+R\leq M\tau+t_{n',0,k'}.
\end{equation}
After the values of $t_{n,m,k}$ satisfying the above conditions are determined, the initial value of $\gamma_{n,m}$ can be constructed by connecting $p_{n,k}$ and $p_{n,k+1}$ with constant velocity motions.

Now, we construct a directed graph such that its vertices are tuples $(n,m,k)$ and its edges connect from $(n,m,k)$ to $(n,m,k+1)$ or from $(n,m,k)$ to $(n',m',k')$ for $(n,m,k,n',m',k')\in C$. Then, for the constraints given in this paper, the graph is acyclic and the indegrees of vertices of the form $(n,m,0)$ are zero. So we can compute the minimum values of $t_{n,m,k}$ satisfying (\ref{eq:zero condition}), (\ref{eq:L condition}), and (\ref{eq:C1 condition}) by dynamic programming after the value of $\tau$ is determined.

Since the value of $\tau$ is not determined yet, we write $t_{n,m,k}=a_{n,m,k}\tau + b_{n,m,k}$ and define the order of pairs of $(a,b)\in \mathbb{Z}\times\mathbb{R}$ by the lexicographical order. Then, we can compute the minimum values of $a_{n,m,k}, b_{n,m,k}$ by dynamic programming satisfying the following conditions:
\begin{equation}
    (a_{n,m,0},b_{n,m,0})=(0,0),
\end{equation}
\begin{equation}
    (a_{n,m,k}.b_{n,m,k}+L_{n,m})\leq (a_{n,m,k+1}, b_{n,m,k+1}),
\end{equation}
\begin{equation}
    (a_{n,m,k}+m, b_{n,m,k}+R) \leq (a_{n',m',k'}+m', b_{n',m',k'})
\end{equation}
for all $(n,m,k,n',m',k')\in C_1$.

We determine the values of $a_{n,m,k}, b_{n,m,k}$ as the minimum ones.
Furthermore, we can prove that these values hold the following inequality by the forms of conditions:
\begin{equation}
    a_{n,m,k}+m\leq M-1.
\end{equation}
Thus, when $\tau$ is large enough, $t_{n,m,k}=a_{n,m,k}\tau+b_{n,m,k}$ while satisfying the conditions (\ref{eq:zero condition}), (\ref{eq:L condition}), (\ref{eq:C1 condition}), and (\ref{eq:C2 condition}). We determine the value of $\tau$ as the minimum one of such values.

\section{C: Implementation Details}
All the methods are implemented in C++ and evaluated with Intel Core i9-9900K CPU and 32 GB RAM.
\paragraph{Optimization method.}
For optimization, we use Levenberg-Marquardt (LM) Algorithm implemented in g$^2$o \cite{kummerle2011g}. Parameters for the optimization are set as the default values of the library.

We set $\sigma_{\mathrm{r}}=\sigma_{\mathrm{v}}=\sigma_{\mathrm{o}}=\sigma_{\mathrm{c}}=10^4$ and $\sigma_{\mathrm{t}}=1$ as the initial values. Since the set $C$ in Eq.(7) of the main paper may change depending on the value of $p$ or $T_{i,j}$, we have to reset the explicit constraints repeatedly during optimization. The number of iteration of LM algorithm and intervals of resetting constraints are as follows:
\begin{enumerate}
\item The first $500$ iterations: we recalculate $C$ after each iteration.
\item The next $39500$ iterations: we recalculate $C$ for $10$ iterations.
\item The following $1000$ iterations: we recalculate $C$
and increase $\sigma_{\mathrm{r}}, \sigma_{\mathrm{v}}, \sigma_{\mathrm{o}}$, and  $\sigma_{\mathrm{c}}$ by multiplying $1.01$ and decrease $\sigma_{\mathrm{t}}$ by dividing $1.01$ after each iteration.
\item The remaining iterations: we recalculate $C$ after each iteration. This phase lasts until the plan converges, more precisely, until the difference of costs for $100$ iterations goes below $10^{-6}$.

\begin{table}[t]
    \centering
    \begin{tabularx}{\linewidth}{l|cccccc}
    \toprule[1pt]
    & \multicolumn{6}{c}{\textbf{Environment}}\\
    $M$&(a)&(b)&(c)&(d)&(e)&(f)\\
    \midrule
         $1$& 44537& 42899& 41113& 45025& 41133 & 46111 \\
         & 2& 4& 5& 13 & 29 &54 \\
    \midrule
         $2$& 74280& 97481& 84898& 49504& 43514& 73906\\
         & 15& 69& 59& 50& 69& 277\\
    \midrule
         $3$& 41910& 83786& 52978& 85698 & 61067& 41973\\
         & 8& 96& 57& 281 & 188& 107\\
    \bottomrule[1pt]
    \end{tabularx}
    \caption{Number of iterations (top) and required time [min] (bottom) for until the optimization converges for the proposed approach.}
    \label{tab:number of iterations}
\end{table}

Table \ref{tab:number of iterations} reports the number of iterations until the optimization converges and required times in minutes.

\end{enumerate}
\subsubsection{Baseline methods.}
For the FCFS approach, we modify the implementation of Prioritized SIPP by \citet{kasaura2022prioritized} to our problem setting. For the SO approach, we modify the implementation of CCBS by the authors, which is available online\footnote{\texttt{github.com/PathPlanning/Continuous-CBS}}, to our problem setting.
For each query, planning was considered a failure if the runtime exceeded $1.0$ seconds.
For PRM, We also use the implementation by \citet{kasaura2022prioritized}. 
As the parameters for PRM,  the number of roadmap vertices is $1000$ for the FCFS approach and $20$ for the SO approach. We limit the number of vertices for the SO because CCBS is generally time-consuming when the density of vertices is high and the environments are small. The number of neighbors is $15$ for the FCFS approach and $10$ for the SO approach.

\clearpage

\bibliography{ref.bib}
\end{document}